\documentclass[reprint,preprintnumbers,amsmath,amssymb,aps,prd,superscriptaddress,nofootinbib]{revtex4-2}

\usepackage{graphicx}% Include figure files
\usepackage{dcolumn}% Align table columns on decimal point
\usepackage{bm}% bold math
\usepackage{hyperref}% add hypertext capabilities
\usepackage[mathlines]{lineno}% Enable numbering of text and display math
\usepackage{xcolor}
\usepackage{booktabs}
\usepackage{enumitem,kantlipsum}

\newcommand{\evSi}{eV$_{\mathrm{Si}}$}
\newcommand{\evHe}{eV$_{\mathrm{He}}$}

\begin{document}

\preprint{}

% \linenumbers

\title{First Demonstration of the HeRALD Superfluid Helium Detector Concept}% Force line breaks with \\

\author{R. Anthony-Petersen} \affiliation{University of California Berkeley, Department of Physics, Berkeley, CA 94720, USA}
\author{A. Biekert} \affiliation{University of California Berkeley, Department of Physics, Berkeley, CA 94720, USA}
\author{C.L. Chang} \affiliation{Argonne National Laboratory, 9700 S Cass Ave, Lemont, IL 60439, USA}
\author{Y. Chang} \affiliation{University of California Berkeley, Department of Physics, Berkeley, CA 94720, USA}
\author{L. Chaplinsky} \affiliation{University of Massachusetts, Amherst Center for Fundamental Interactions and Department of Physics, Amherst, MA 01003-9337 USA}
\author{A. Dushkin} \affiliation{University of Michigan, Randall Laboratory of Physics, Ann Arbor, MI 48109-1040, USA}
\author{C.W. Fink} \affiliation{University of California Berkeley, Department of Physics, Berkeley, CA 94720, USA}
\author{M. Garcia-Sciveres} \affiliation{Lawrence Berkeley National Laboratory, 1 Cyclotron Rd., Berkeley, CA 94720, USA} \affiliation{International Center for Quantum-field Measurement Systems for Studies of the Universe and Particles (QUP,WPI), High Energy Accelerator Research Organization (KEK), Oho 1-1, Tsukuba, Ibaraki 305-0801, Japan}
\author{W. Guo} \affiliation{Department of Mechanical Engineering, FAMU-FSU College of Engineering, Florida State University, Tallahassee, FL 32310, USA} \affiliation{National High Magnetic Field Laboratory, Tallahassee, FL 32310, USA}
\author{S.A. Hertel} \affiliation{University of Massachusetts, Amherst Center for Fundamental Interactions and Department of Physics, Amherst, MA 01003-9337 USA}
\author{X. Li} \affiliation{Lawrence Berkeley National Laboratory, 1 Cyclotron Rd., Berkeley, CA 94720, USA}
\author{J. Lin} \affiliation{University of California Berkeley, Department of Physics, Berkeley, CA 94720, USA}
\author{R. Mahapatra} \affiliation{Texas A\&M University, Department of Physics and Astronomy, College Station, TX 77843-4242, USA}
\author{W. Matava} \affiliation{University of California Berkeley, Department of Physics, Berkeley, CA 94720, USA}
\author{D.N. McKinsey} \affiliation{University of California Berkeley, Department of Physics, Berkeley, CA 94720, USA} \affiliation{Lawrence Berkeley National Laboratory, 1 Cyclotron Rd., Berkeley, CA 94720, USA}
\author{D.Z. Osterman} \affiliation{University of Massachusetts, Amherst Center for Fundamental Interactions and Department of Physics, Amherst, MA 01003-9337 USA}
\author{P.K. Patel} \affiliation{University of Massachusetts, Amherst Center for Fundamental Interactions and Department of Physics, Amherst, MA 01003-9337 USA}
\author{B. Penning} \affiliation{University of Michigan, Randall Laboratory of Physics, Ann Arbor, MI 48109-1040, USA}
\author{H.D. Pinckney} \thanks{Corresponding author: \href{mailto:hpinckney@umass.edu}{hpinckney@umass.edu}}\affiliation{University of Massachusetts, Amherst Center for Fundamental Interactions and Department of Physics, Amherst, MA 01003-9337 USA}
\author{M. Platt} \affiliation{Texas A\&M University, Department of Physics and Astronomy, College Station, TX 77843-4242, USA}
\author{M. Pyle} \affiliation{University of California Berkeley, Department of Physics, Berkeley, CA 94720, USA}
\author{Y. Qi} \affiliation{Department of Mechanical Engineering, FAMU-FSU College of Engineering, Florida State University, Tallahassee, FL 32310, USA} \affiliation{National High Magnetic Field Laboratory, Tallahassee, FL 32310, USA}
\author{M. Reed} \affiliation{University of California Berkeley, Department of Physics, Berkeley, CA 94720, USA}
\author{G.R.C Rischbieter} \affiliation{University of Michigan, Randall Laboratory of Physics, Ann Arbor, MI 48109-1040, USA}
\author{R.K. Romani} \affiliation{University of California Berkeley, Department of Physics, Berkeley, CA 94720, USA}
\author{A. Serafin} \affiliation{University of Massachusetts, Amherst Center for Fundamental Interactions and Department of Physics, Amherst, MA 01003-9337 USA}
\author{B. Serfass} \affiliation{University of California Berkeley, Department of Physics, Berkeley, CA 94720, USA}
\author{R.J.  Smith} \affiliation{University of California Berkeley, Department of Physics, Berkeley, CA 94720, USA}
\author{P. Sorensen} \affiliation{Lawrence Berkeley National Laboratory, 1 Cyclotron Rd., Berkeley, CA 94720, USA}
\author{B. Suerfu} \affiliation{International Center for Quantum-field Measurement Systems for Studies of the Universe and Particles (QUP,WPI), High Energy Accelerator Research Organization (KEK), Oho 1-1, Tsukuba, Ibaraki 305-0801, Japan}
\author{A. Suzuki} \affiliation{Lawrence Berkeley National Laboratory, 1 Cyclotron Rd., Berkeley, CA 94720, USA}
\author{V. Velan} \affiliation{Lawrence Berkeley National Laboratory, 1 Cyclotron Rd., Berkeley, CA 94720, USA}
\author{G. Wang} \affiliation{Argonne National Laboratory, 9700 S Cass Ave, Lemont, IL 60439, USA}
\author{Y. Wang} \affiliation{University of California Berkeley, Department of Physics, Berkeley, CA 94720, USA}
\author{S.L. Watkins} \affiliation{University of California Berkeley, Department of Physics, Berkeley, CA 94720, USA}
\author{M.R. Williams} \affiliation{University of Michigan, Randall Laboratory of Physics, Ann Arbor, MI 48109-1040, USA}

\collaboration{SPICE/HeRALD Collaboration}%\noaffiliation

\date{\today}% It is always \today, today,
             %  but any date may be explicitly specified

\begin{abstract}

The SPICE/HeRALD collaboration is performing R\&D to enable studies of sub-GeV dark matter models using a variety of target materials.  Here we report our recent progress on instrumenting a superfluid $^4$He target mass with a transition-edge sensor based calorimeter to detect both atomic signals (scintillation) and $^4$He quasiparticle (phonon and roton) excitations.  The sensitivity of HeRALD to the critical ``quantum evaporation'' signal from $^4$He quasiparticles requires us to block the superfluid film flow to the calorimeter.  We have developed a heat-free film-blocking method employing an unoxidized Cs film, which we implemented in a prototype ``HeRALD v0.1'' detector of $\sim$10~g target mass.  This article reports initial studies of the atomic and quasiparticle signal channels.  A key result of this work is the measurement of the quantum evaporation channel's gain of $0.15 \pm 0.01$, which will enable $^4$He-based dark matter experiments in the near term.  With this gain the HeRALD detector reported here has an energy threshold of 145~eV at 5 sigma, which would be sensitive to dark matter masses down to 220~MeV/c$^2$.

\end{abstract}

\maketitle

\section{\label{sec:level1} Introduction}

Dark matter (DM) models in which the DM mass is between 1~keV/c$^2$ and 1~GeV/c$^2$ are sometimes termed `low-mass DM', in contrast to DM models based on new physics at the $>$1~GeV/c$^2$ electroweak scale.  Low-mass DM has received increasing attention over the past decade as it has become clear that these models are cosmologically viable, are only loosely constrained by existing experiments, and retain comparative model simplicity.  Such DM models include the ELDER~\cite{kuflikElasticallyDecouplingDark2016,kuflikPhenomenologyELDERDark2017a}, SIMP~\cite{hochbergMechanismThermalRelic2014,hochbergModelThermalRelic2015}, and `freeze-in'~\cite{hallFreezeinProductionFIMP2010,elor_maximizing_2023} frameworks.  These models are testable through direct detection, either via electron scattering or nuclear scattering processes.  In the nuclear recoil case, deposited energy scales with the square of the DM mass, and these models require highly sensitive detectors to study.  To set the energy scale, even for a DM mass relatively high in the light-DM range (10~MeV/c$^2$) and the relatively low-mass target nucleus of $^4$He, the recoil endpoint energy will be $\sim$100~meV.  The TESSERACT (Transition-Edge Sensors with Sub-eV Resolution And Cryogenic Targets) R\&D program~\cite{d.mckinseySnowmass2021Letter} is developing new sensor technologies to measure these sub-eV recoil energy signatures by advancing Transition-Edge Sensors (TESs) to sub-eV phonon-sensing thresholds and applying them to a variety of target materials with diverse and complementary DM sensitivities.  The SPICE portion of the TESSERACT program uses substrates with strong optical phonon modes, while `HeRALD' (Helium Roton Apparatus for Light Dark matter)~\cite{hertelDirectDetectionSubGeV2019} uses superfluid $^4$He.

%These sub-eV recoil energy signatures motivate the development of new sensor technologies which is at the core of the TESSERACT R\&D program, in which Transition Edge Sensors (TESs) are being advanced to sub-eV phonon-sensing thresholds, and are being applied to a variety of target materials with diverse and complementary DM sensitivities.  The $^4$He portion of the TESSERACT program is termed `HeRALD' (Helium Roton Apparatus for Light Dark matter)~\cite{d.mckinseySnowmass2021Letter,hertelDirectDetectionSubGeV2019}.

\subsection{Superfluid $^4$He as a target material}

Superfluid $^4$He offers several advantages as a DM target material. First, the nuclear mass is comparatively low, which correspondingly boosts the nuclear recoil spectrum endpoint energy.  The recoil endpoint scales inversely with the atomic mass A, meaning $^4$He will expect a $\sim$7$\times$ ($\sim$33$\times$) increase in endpoint energy relative to Si (Xe).

We also expect the backgrounds in the sub-eV range to be low compared to other technologies, such as solid-state detectors.  As all impurities freeze out to the boundaries (\cite{sprague_3mathrmhe_1994,akimoto_coverage_2006} and references within), $^4$He is extremely radiopure at mK temperatures.  Additionally, while crystal lattices typically contain dislocations and stress which can lead to spontaneous phonon emission even at zero temperature~\cite{adariEXCESSWorkshopDescriptions2022,anthony-petersenStressInducedSource2022,astromFractureProcessesObserved2006,angloherLatestObservationsLow2022}, a superfluid material should minimize such stored potential energies and the spontaneous phonon backgrounds arising from them.  Further, the first electronic excited state of $^4$He requires 19.8~eV.  Therefore, Compton scattering processes are entirely forbidden in the wide 0-19.8~eV energy range key to low-mass DM sensitivity.  The dominant background in this eV range will instead be coherent gamma scattering (see Reference \cite{robinson}). 

Yet another advantage is superfluid $^4$He's complement of three unique signal channels: quasiparticle-induced ``quantum evaporation''~\cite{bairdQuantizedEvaporationLiquid1983, bandlerParticleDetectionEvaporation1992}, and electronically-excited dimers in their singlet and triplet states~\cite{danielnicholasmckinseyDetectionMagneticallyTrapped2002, mckinseyRadiativeDecayMetastable1999, mckinseyTimeDependenceLiquidhelium2003}.  In this context,  ``quasiparticle'' describes a range of phonon-like excitations of the superfluid medium, ranging from low-momentum phonons to higher-momentum R$^-$ and R$^+$ ``rotons''.  The quantum evaporation signal is produced when single $^4$He quasiparticles liberate single $^4$He atoms from the liquid surface in a one-to-one process, sending them into the vacuum above.  While this process has an 0.62~meV threshold, if the liberated atoms are sensed calorimetrically then the signal energy is not the atoms' comparatively small kinetic energy, but instead the much larger van der Waals potential of $^4$He adsorption at the calorimeter surface: roughly 10~meV per atom~\cite{vidaliPotentialsPhysicalAdsorption1991}.

Particle interactions greater than 19.8~eV also produce electronic excitations.  These excitations lead to the formation of He$_2^*$ dimers which (as in other liquid noble elements) appear in either a short-lived singlet state ($<$10~ns) or a long-lived triplet state (13~s)~\cite{danielnicholasmckinseyDetectionMagneticallyTrapped2002,mckinseyTimeDependenceLiquidhelium2003,mckinseyRadiativeDecayMetastable1999}.  While the singlet excimers promptly decay to emit $\sim$15.5~eV VUV photons, the triplet excimers decay through one of two processes.  First, if two triplet excimers collide their combined internal energy is sufficient to ionize one of the atoms (Penning ionization).  Following this ionization a new excimer will form: if it is in the singlet state it will decay promptly, otherwise it re-enters the triplet population.  As the density of triplets is initially high at the recoil site some triplets contribute to the prompt scintillation signal.  Triplets that escape the recoil site propagate ballistically through the bulk superfluid.  Here they can once again de-excite through collisions with each other, and they can also de-excite at surfaces~\cite{carter_calorimetric_2017}.  For more detail on these signal channels and the HeRALD concept, see Ref.~\cite{hertelDirectDetectionSubGeV2019}.  Combining this scintillation channel with the quasiparticle channel could provide event-by-event information on the electron or nuclear recoil nature of interactions, though this becomes irrelevant below 19.8~eV as electron recoils are forbidden.

The dual-signal readout of $^4$He in both quasiparticles (via quantum evaporation) and scintillation has been previously explored experimentally by the HERON collaboration \cite{lanouDetectionSolarNeutrinos1987,bandlerParticleDetectionEvaporation1992}.  While the original motivation of the HERON program was the detection of pp solar neutrinos via keV-scale electron recoils, the application to DM direct detection via nuclear recoils was also considered~\cite{adamsHERONDarkMatter1996}.

Finally, $^4$He carries two notable disadvantages as a target material.  While the light nuclear mass increases energy sensitivity, it decreases the cross-section for spin-independent interactions, assumed to scale as A$^2$.  Additionally, as $^4$He is a spin-zero atom it will have no sensitivity to spin-dependent interactions.

\subsection{Superfluid film-stopping}\label{sec:film_stopping}

Though $^4$He offers numerous advantages as an active medium its superfluid nature also poses practical challenges~\cite{frankpobellMatterMethodsLow2007}.  First, the target and sensors must be contained within a superfluid-tight volume.  More challengingly, a superfluid $^4$He film typically covers all surfaces within such a closed container~\cite{vidaliPotentialsPhysicalAdsorption1991,bairdQuantizedEvaporationLiquid1983}.  The presence of a $^4$He film on a calorimetric sensor would significantly degrade the sensor's performance, first by adding additional heat capacity and energy escape mechanisms, and second by negating the highly advantageous van der Waals gain mechanism.  The HERON collaboration prevented film flow to the sensor by implementing a ``film-burner'', in which a region of the cell is heated, interrupting the film flow to the calorimeter via evaporation~\cite{toriiRemovalSuperfluidHelium1992}.  While successfully demonstrated, the film burner strategy introduces an unavoidable heatload to the calorimeter itself, raising the sensor's energy threshold. As energy threshold is the main design driver for the HeRALD program, we investigated alternative \emph{heat-free} technologies for film stopping.  One method, the ``knife-edge'' method, has been well demonstrated at higher temperatures, but it was found to be ineffective in our experimental configuration (see Appendix~\ref{sec:cryostat}).

It has been  predicted~\cite{chengHeliumPrewettingNonwetting1991} and subsequently demonstrated~\cite{ketolaAnomalousWettingHelium1992,nacherExperimentalEvidenceNonwetting1991,rutledgePrewettingPhaseDiagram1992} that superfluid $^4$He films do not wet unoxidized cesium.  Non-wetting by $^4$He is an extremely rare material property, specific only to the heaviest alkali metals (perhaps only cesium and rubidium~\cite{chengHeliumPrewettingNonwetting1991}).  To implement a cesium-based film-stopping method we deposit a cesium film on a region of the cell between the superfluid target and the sensor used for scintillation and evaporated atoms.  To avoid oxidation, this cesium deposition must occur \emph{in situ} within the detector during the cool-down.  Implementation of this approach is described in Sec.~\ref{sec:operation} and Appendices~\ref{sec:cryostat} and \ref{sec:detector_preparation}.

\subsection{HeRALD \MakeLowercase{v}0.1}

This article summarizes work with a prototype HeRALD detector, which we term v0.1.  The goal of this prototype was to observe particle interactions in a superfluid $^4$He target using a single sensor, suspended in the vacuum over the target and kept dry by a cesium-based film stopper.

\section{Detector Setup, Cesium Testing, and Data collection}\label{sec:operation}

\subsection{The HeRALD \MakeLowercase{v}0.1 Detector}\label{sec:detector_main}

\begin{figure}[tbp]
    \begin{center}
        \includegraphics[width=1.\columnwidth]{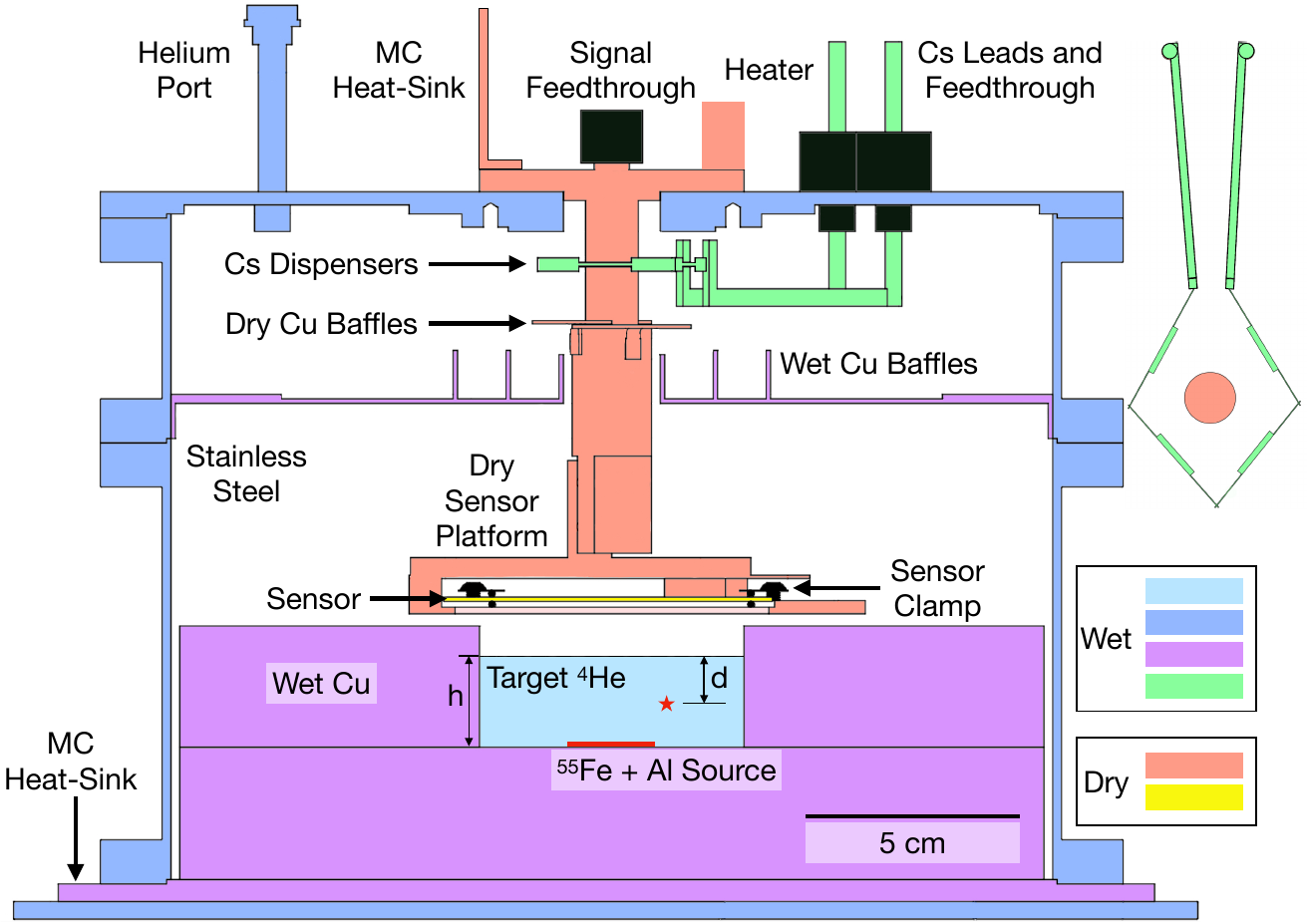}
    \end{center}
    \caption{Annotated cross section of the detector used in this work.  The $^4$He-wetted surfaces are the stainless steel cell boundaries (blue), the copper shielding plates (purple), and the cesium dispensers (green).  The $^4$He-free ``dry'' surfaces are suspended from the cesium region:  the sensor platform (orange), the silicon ``3-inch-diameter-standard'' calorimetric sensor itself (yellow), and the sapphire ball detector clamps (light blue). The target $^4$He volume is represented as light blue.  As referenced in the text, the total $^4$He fill height ``$h$'' and the depth of an interaction ``$d$'' are defined as indicated.  An X-ray source at the bottom of the $^4$He target region is also indicated.  A top-down view of the cesium evaporator configuration is to the right, along with a color legend.}
    \label{cell_drawing}
\end{figure}

To accomplish the goals of HeRALD v0.1 we assembled the detector system illustrated in Fig.~\ref{cell_drawing}, referred to as the ``cell''.  This cell is accompanied by a number of cryostat systems described in Appendix~\ref{sec:cryostat}.   

At the heart of the detector system is the $^4$He-free (dry) ``sensor platform'' which supports a large-area calorimeter above the target $^4$He volume.  This calorimeter is an iteration of the ``Cryogenic Photon Detector''~\cite{finkPerformanceLargeArea2021, supercdmscollaborationLightDarkMatter2021}, based on a 3-inch-diameter-standard Si wafer of 1~mm thickness.  The upper surface of the wafer is instrumented to sense athermal Si phonons using an array of tungsten transition-edge sensors (TESs) with a critical temperature of 51~mK.  Athermal phonons in the silicon substrate are absorbed by aluminum fins, which transport the energy into the TESs in a standard ``QET'' architecture~\cite{irwinQuasiparticleTrapAssisted1995}.  The 673 TESs are connected in parallel as a single channel, and throughout the text we will refer to the calorimeter as the ``sensor''.  After fabrication, a single TES was disconnected from the array and repurposed as a resistive heater in order to apply heat to the sensor directly.  The sensor is mechanically secured to the sensor platform using three sapphire-ball-based clamps~\cite{pinckneyThermalConductanceSapphire2022}.  In each of the three clamps, the wafer is pressed between two 1.5~mm diameter sapphire balls with one hard-temper phosphor bronze spring.  A weak thermal link between the calorimeter and the platform is constructed using a gold pad and a gold wirebond. The TES array and heater are connected electrically via aluminum wire bonds to superconducting PCBs on the platform, which in turn connect to NbTi twisted pairs.  These twisted pairs exit the experimental cell through a homemade Stycast 2850FT feedthrough.  The sensor platform has a dedicated thermal link to the mixing chamber stage consisting of two copper (grade 101) sheets in parallel, each being 3~cm wide $\times$ 10~cm long $\times$ 0.5~mm thick.  A thermometer and resistive heater were secured to the sensor platform structure on the outside of the cell.  

Directly below the sensor platform is the $^4$He target region: a 2.75~cm high, 6~cm diameter cylinder.  This region is defined by two copper plates which also provide some shielding of the region from external gamma backgrounds.  We will refer to the amount of $^4$He in this region by its fill height ``$h$'', and separately the depth of an interaction relative to the $^4$He surface as ``$d$''.  The $^4$He enters this target volume through a 1.7~mm ID, 2.1~mm OD stainless steel capillary connected via a VCR flange on the top of the cell.  

The calibration source is immersed within the target $^4$He volume and consists of $^{55}$Fe on a thin copper substrate, laminated with $\sim$75~$\mu$m of polypropylene, and covered with an $\sim$25~$\mu$m thick layer of Al.  The dominant X-ray produced by $^{55}$Fe is 5.9~keV.  We expect a 5.9~keV electron-recoil-like event to deposit 24\% (1400~eV), 32\% (1875~eV), and 34\% (1975~eV) of the recoil energy into the triplet excimer, singlet excimer, and quasiparticle channels of the detector respectively, with the remainder (10\%) appearing as infrared radiation.  These partitions were estimated using a simulation code which implements the scintillation yields from~\cite{collaborationScintillationYieldElectronic2022} and is inspired by the \texttt{NEST} package~\cite{szydagis_m_2023_7577399,szydagisReviewNESTModels2023}.  A higher-energy X-ray at 6.5~keV is also produced, but at lower flux (lower by a ratio of 1:7.3). The Al layer is added specifically to produce a useful fluorescence X-ray at 1.5~keV.  The source activity was $\sim$140~Bq, corresponding to a rate of $\mathcal{O}(100)$ $^{55}$Fe events per hour in the target $^4$He.  In addition to providing calibration peaks at 5.9 and 1.5~keV, these two X-ray energies exhibit distinct path lengths in the superfluid: 0.4~cm at 1.5~keV and 25~cm at 5.9~keV.  This difference in path length allows us to have two different populations under study: one concentrated at the bottom center of the $^4$He, and another more uniformly distributed throughout the target.  

The cesium-based film stopping system occupies the upper region of the cell as illustrated in Fig.~\ref{cell_drawing}.  Four commercial cesium dispensers\footnote{CS/NF/3.9/12 FT10+10 from SAES} evaporate a Cs vapor when heated using several Amps of current.  These dispensers are arranged in a ring and are mechanically supported only by their two copper current leads, which enter the cell via a second homemade Stycast 2850FT feedthrough.  To avoid Cs deposition on the sensor itself, copper baffling prevents any line-of-sight path between the dispensers and the sensor.

All of these systems are contained within a surrounding stainless steel experimental cell. The cell is sealed using indium and is supported structurally from its base.  The structural support also serves as a thermal link between a copper base plate (which forms a portion of the cell interior) and the mixing chamber stage.

\subsection{Cesium Efficacy}\label{sec:cesium}

The deposition of the cesium film begins with the cell surfaces at $\sim$4~K, ensuring low ambient pressure due to the cryogenic vacuum.  A current is applied to the commercial cesium dispensers, ramping up to 7.5~A (a dispenser temperature of $\sim800~^{\circ}$C).  This heats the sensor portion of the cell to $\sim$70~K.  After deposition, the cesium current leads are mechanically disconnected within the vacuum space before the cooldown continues to a $\sim$7~mK base temperature.  More details are presented in Appendix~\ref{sec:detector_preparation}.

While at base temperature but before condensing $^4$He within the cell, the calorimeter is characterized in a $^4$He-free state.  This provides us with a sensor calibration based on the 1.5 keV Al fluorescence X-ray (Section~\ref{sec:calibration}), and a reference point of sensor response in the $^4$He-free state.  Condensation then proceeds via a 20~sccm flow from room temperature within the capillary.  While the majority of $^4$He arrives at the detector already condensed within the capillary as a superfluid, it is observed that some $^4$He arrives at the detector in a gas phase, thereby wetting the sensor platform region and necessitating a post-fill bake of the sensor platform to remove this film.  Short-duration (2~second) heat pulses are applied first to the sensor platform and then directly to the Si wafer through the TES heater removed from the array. The heat pulse to the platform raises both the platform and silicon to 1~K, and the heat pulse to the silicon raises the Si temperature to $\sim$1~K while keeping the platform below 500~mK (Si temperature estimated through thermal modeling).  This high Si temperature is required in order to remove the last atomic layer of $^4$He, and is achievable due to the engineered thermal conductance between the sensor and mixing chamber.  If the cesium system is working, we should see that no $^4$He re-enters the sensor platform following these heat pulses.

\begin{figure}[htbp]
    \begin{center}
        \includegraphics[width=1.\columnwidth]{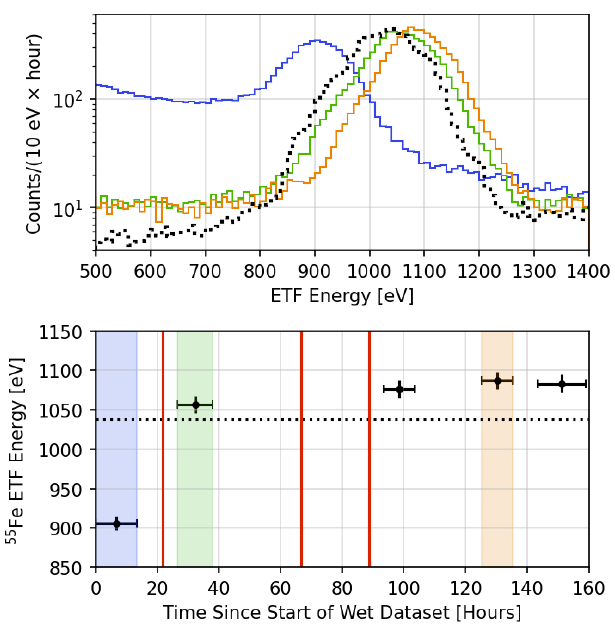}
    \end{center}
    \caption{A sensor baking procedure to remove the $^4$He film, demonstrating the efficacy of the cesium-based superfluid-film blocking system.  \textit{Top:} four spectra in the region of the 5.9~keV X-ray peak from $^{55}$Fe: (black dotted) no $^4$He, (blue) after filling $^4$He, (green) after initial bakes, and (orange) after additional bakes.  \textit{Bottom:} The reconstructed peak energy of the $^{55}$Fe X-ray source interacting with the silicon sensor, where the peak energy is computed with the electrothermal feedback (ETF) integral method from~\cite{irwinTransitionEdgeSensors2005}.  Vertical red lines indicate detector heating. Shaded color bars highlight the time period for the later three spectra.  The horizontal black dotted line indicates the dry state peak height.  Horizontal error bars represent the span of the data set used to compute the peak amplitude, and error bars in energy represent an uncertainty in bias point.}
    \label{fig:cs}
\end{figure}

The effect of this $^4$He removal procedure is shown in Fig.~\ref{fig:cs}, where the times of three heat-pulse series are indicated with vertical red lines on the bottom panel.  The amount of $^4$He film present can be estimated using the Si phonon transport efficiency, monitored using a calibration X-ray peak at 5.9~keV. As the TES current is related to the power it dissipates, the total energy dissipated during an event can be inferred from the pulse integrals (the electrothermal feedback integral method of Ref.~\cite{irwinTransitionEdgeSensors2005}).  In Fig.~\ref{fig:cs} we use this technique to monitor the 5.9~keV signal response over the course of this example baking procedure.  The upper panel shows the 5.9~keV spectral peak as it appears in the detector before filling (meaning no $^4$He), after filling, and after baking.  The lower panel tracks the 5.9~keV peak signal amplitude throughout the filling and baking procedure.  Horizontal ranges indicate the time window used to estimate the 5.9~keV response, and error bars in the energy axis represent an uncertainty in TES bias.  Following the $^4$He fill, the signal amplitude is noticeably suppressed.  Baking steps are seen to increase the signal amplitudes to an asymptotic higher value, in fact exceeding that of the earlier dry state (when the signal was slightly suppressed due to a slightly higher Si temperature).  We see this new higher signal amplitude remains constant over at least a days-long timescale.  We interpret this to mean that the cesium film has succeeded in blocking the flow of superfluid $^4$He to the sensor.

\subsection{$^4$He Data Collection}

The data presented here includes measurements at several $^4$He fill states.  After each $^4$He fill and subsequent sensor bake, the sensor cools sufficiently for data-taking within a 1~h timescale.  We recorded data at six fill heights: \textit{h}~= 9~mm, 13~mm, 19~mm, 22~mm, 25~mm, 27.7~mm.  These fill heights are determined using the pressure drop in the room temperature $^4$He storage tank and the measured detector dimensions.  The final fill height is estimated to be $\sim$0.2~mm over-full (filled over the top of the copper plate, but not in contact with the sensor platform) and showed a sudden increase in low-energy event rate (presumably due to the new large area of $^4$He with a poor geometry for signal collection).  The fill height uncertainty is approximately $\pm2$~mm.

Sensor waveform data was shaped by a 500~kHz low-pass anti-aliasing filter and recorded as a continuous datastream at 1.25~MHz, see Appendix~\ref{sec:cryostat}.  Event finding occurred in subsequent offline analysis (described in Sec.~\ref{sec:processing}).  Data was recorded at each fill height for $\mathcal{O}$(10~h).  Beginning with the \textit{h}~=~22~mm data set the cryostat was surrounded with a water tank of $\mathcal{O}$(10~cm) thickness to further reduce the rate of background Compton scattering by a factor of 2\footnote{Using the ETF energy estimator (Section~\ref{sec:cesium}) the rate between 2 and 3~keV$_{\mathrm{Si}}$ was reduced from $409\pm4$~counts/(keV$_{\mathrm{Si}}$-hour) to $194\pm3$~counts/(keV$_{\mathrm{Si}}$-hour)}.

\section{Data Processing, Data Selection, and Calibration}\label{sec:processing}

\subsection{Particle Interactions in $^4$He and Data Processing}

\begin{figure}[tbp]
    \begin{center}
        \includegraphics[width=\columnwidth]{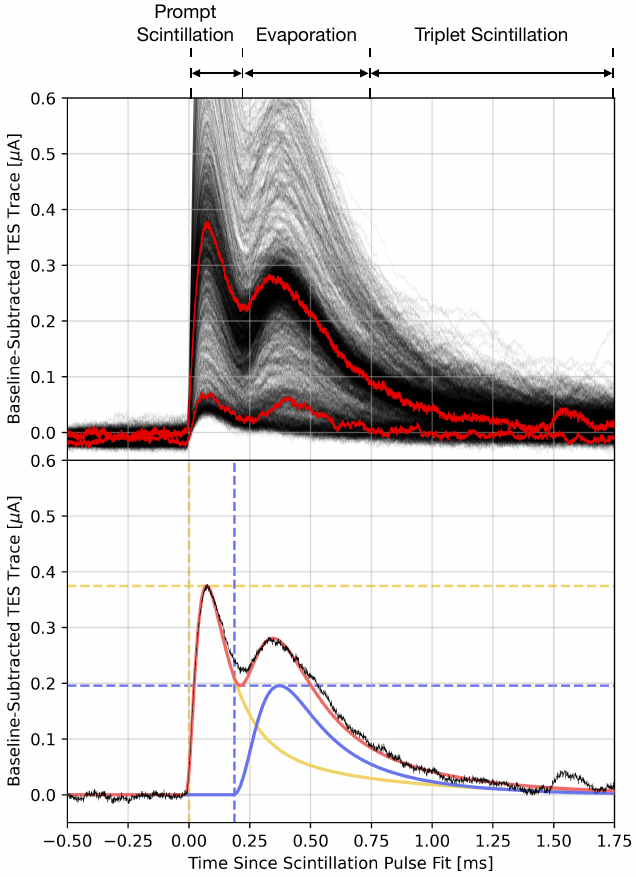}
    \end{center}
    \caption{Annotated waveforms from the $h$~=~13~mm data set (baseline subtracted). \textit{Top}: Overlay of many waveforms.  Scintillation, then evaporation, then triplet-dominated time windows are visible.  The X-ray calibration features at 5.9 and 1.5~keV are visible as denser bands (with evaporation amplitudes at $\sim0.27$~$\mu$A and $\sim0.08$~$\mu$A).  Typical waveforms for both the 5.9 and 1.5~keV calibration lines are highlighted in red.  Waveforms for this plot are selected using the standard quality cuts described in Sec.~\ref{sec:processing}, together with a 40~eV energy threshold on the scintillation pulse.  We also enforce that the event occurs in the $^4$He by requiring an evaporation delay in the observed range of 150-250~$\mu$s, general $\chi^2 < 5\times10^5$, and $\Delta\chi^2$ greater than 0.041 times the evaporation energy squared.  \textit{Bottom}: An example fit using the higher-energy highlighted waveform.  Data is in black, the scintillation and evaporation fits are in yellow and blue respectively, and their sum is shown in red.  The fitted start times and amplitudes are indicated with vertical and horizontal dashed lines respectively, with color indicating whether the quantity belongs to scintillation.}  
    \label{fig:pulses}
\end{figure} 

The several signal channels of $^4$He are separated in arrival time at the calorimeter. Figure~\ref{fig:pulses} shows many $^4$He events from one fill state (h~=~13~mm).  We identify three time intervals within each waveform:
\begin{enumerate}[wide, labelwidth=!, labelindent=0pt]
    \item Prompt scintillation arrives nearly instantaneously with the recoil itself (t~=~0~ms) with a pulse rise and fall determined by the Si phonon collection within the calorimeter.
    \item Quantum evaporation arrives over a broader time after a delay of $\mathcal{O}$(100~$\mu$s).
    \item The quenching of the triplet molecules at $^4$He interfaces dominates the waveform at late times (t$>$0.75~ms), although some level of triplet quenching within the recoil track likely occurs earlier.
\end{enumerate}

Events are analyzed using an optimum filter (OF) framework applied to the data in two steps.  First, events are found using a single-pulse OF, utilizing the scintillation pulse shape as the signal template.  The event finding procedure saves traces of 5~ms length.  After event finding, each waveform is passed through a two-pulse OF which simultaneously identifies the amplitude of the scintillation and evaporation pulses as well as their start times.  The key outputs of this algorithm are a fitted prompt scintillation amplitude, a fitted evaporation amplitude, and the fitted time difference between the two pulses.

For construction of the OF, we estimate the power spectral density of the noise using traces at randomly selected times.  The purity of this noise sample was improved using selection criteria on the baseline average, baseline slope, baseline standard deviation, ``general'' (single-pulse) $\chi^2$, integral-estimated energy, and a pileup event amplitude.  The scintillation pulse template for the OF was assembled by ``stitching'' two pieces:  the rising portion of the pulse was taken directly from averaged $^4$He scintillation waveforms, while the falling portion was taken from events where energy was deposited directly into the Si of the calorimeter.  This stitching of rising and falling portions was performed in order to capture any true effect of $^4$He scintillation light timing at early times, and to avoid contamination at later times from the overlapping evaporation signal.  This scintillation shape was observed to be consistent between data sets.

\begin{figure}[tbp]
    \begin{center}
        \includegraphics[width=\columnwidth]{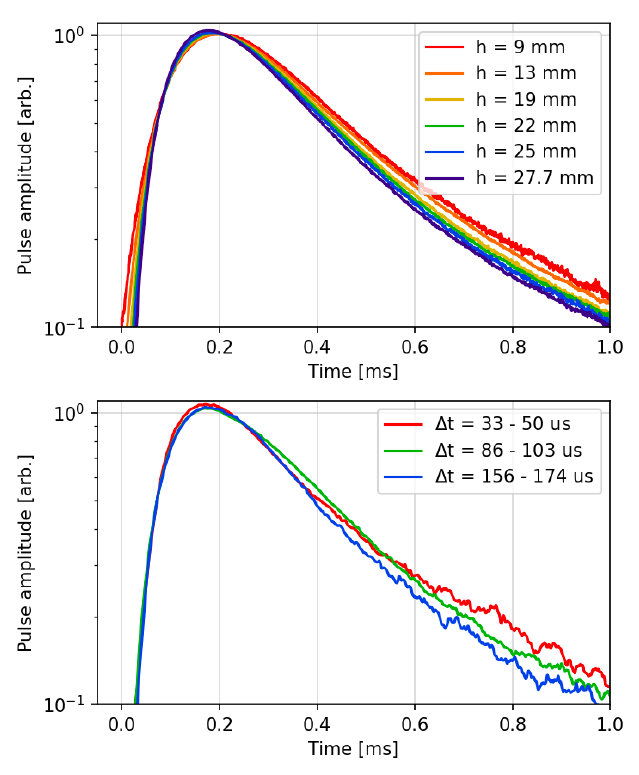}
    \end{center}
    \caption{Residual evaporation pulse shapes after subtracting off the scintillation component (procedure details in Section~\ref{sec:evap_shapes}).  \textit{Top:} median pulse shapes for a variety of fill heights, considering the entire experimental volume.  \textit{Bottom:} partial-volume median pulses (top/middle/bottom of the $^4$He) at one fixed fill height ($h$~=~27.7~mm).  The evaporation shape is observed to change as a function of both ``\textit{h}'' and ``\textit{d}''.  See Section~\ref{sec:time_delay} for a description of how ``\textit{d}'' relates to $\Delta$t. In each panel, \textit{t}~=~0 marks the start of the evaporation pulse as determined by the optimal filter.}
    \label{fig:delay_shapes}
\end{figure}

With the scintillation pulse template now constructed, the evaporation pulse template for the OF was constructed from residuals after scintillation-only fits, with each evaporation-dominated residual waveform shifted according to a fit for the evaporation pulse start time.  We find that evaporation pulse shapes show some level of variation, shown in Fig.~\ref{fig:delay_shapes} and discussed in Sec.~\ref{sec:evap_shapes}.  For simplicity in the formation of an OF for amplitude and delay time estimation, we choose a single ``typical'' evaporation waveform (the average evaporation waveform for \textit{h}~=~22~mm data) to play the role of the evaporation signal template in the OF processing of all data.  The slight mismatch between this ``typical'' evaporation waveform and any specific waveform introduces some systematic error in the OF amplitude estimator.  We quantify this systematic error by applying a varied choice of OF template waveform to the same data, and find that the systematic error on estimated evaporation amplitude is less than $\sim$5\% throughout this analysis.

\subsection{Data Selection}\label{sec:data_selection}

Throughout this work a number of ``standard quality cuts'' are applied to the data in order to ensure that we are measuring interactions in a quiescent detector.  Within each dataset these quality cuts sequentially remove:

\begin{enumerate}[wide, labelwidth=!, labelindent=0pt]
    \item Events whose pre-pulse baseline average is greater than the minimum baseline in the dataset plus 1\%
    \item Events whose baseline slope is further than two standard deviations from the dataset mean
    \item Events whose  baseline standard deviation falls below the dataset mode plus the difference between the dataset mode and minimum
\end{enumerate}

To select events specifically within the $^4$He (as opposed to the Si), we apply another series of cuts referred to as ``in-$^4$He''.  First we select in evaporation time delay, see Table~\ref{tab:delay_cuts}.  This selection is conservatively loose, above and below the clear ``interactions in $^4$He'' band for each fill level (for reference, see Figs.~\ref{fig:delay_times} and ~\ref{fig:amplitudes_vs_delay}).  Additionally, unless otherwise stated ``general $\chi^2$'' refers to a $\chi^2$ selection using a single-template OF fit ($\chi^2 < 10^6$), and ``$\Delta\chi^2$'' refers to a selection based on the difference in $\chi^2$ between a two-template and single-template OF fit ($\Delta\chi^2 > 100$).  Each of these selections removes electronic noise events and helps to ensure we are only considering interactions in the target region.  The $\Delta\chi^2$-based distinction between Si `direct hits' and $^4$He `double pulses' becomes challenging near threshold, where we approach the noise floor of the sensor and also the single-detected-photon limit of the scintillation signal.  We also enforce that the fit of the evaporation amplitude is positive.  

\begin{table}[htb]
    \begin{center}
        \begin{tabular}{lll}
        \toprule
        \textit{h} [mm] \hspace{5mm} & Minimum [$\mu$s] \hspace{5mm} & Maximum [$\mu$s]\\
        \midrule
        9    & 200    &  300\\
        13   & 150    &  235\\
        19   & 100    &  230\\
        22   & 75     &  230\\
        25   & 50     &  230\\
        27.7 & 15     &  230\\
        \bottomrule
        \end{tabular}
    \end{center}
    \caption{Evaporation time delay cut values for each fill height ``\textit{h}''.}
    \label{tab:delay_cuts}
\end{table}

In certain analyses we also apply an amplitude selection for $^{55}$Fe events in $^4$He, referred to as the ``$^{55}$Fe'' cut.  This selection is trapezoidal in the space of delay time vs. prompt scintillation energy of Fig.~\ref{fig:amplitudes_vs_delay}, and is again defined to be conservatively loose, to ensure inclusion of the full $^{55}$Fe distribution.

\subsection{Calibration}\label{sec:calibration}

The detector energy scale is calibrated using the 1.5~keV Al fluorescence peak appearing directly in Si in pre-$^4$He data.  For the specific analyses of this work, the calibration data was obtained from a previous cooldown of the experiment, leading to a 6\% systematic uncertainty on this calibration (estimated using the lower-rate 1.5~keV flux reaching the Si at the lowest $^4$He fill state of this cooldown).  Using the 1.5~keV peak, we compare the energy reconstructed using the electrothermal feedback method (Section~\ref{sec:cesium} and~\cite{irwinTransitionEdgeSensors2005}) and the known energy of the X-ray peak to measure the transport efficiency of Si phonon energy into the TES system, 26\%.  We assume a linear calibration between this 1.5~keV energy and 0~eV, informed by validating the pulse shape does not significantly change throughout this energy range.  

An additional correction is made to account for detector state variation throughout the multiple-week data-taking period.  The reference point for these corrections is the $^{55}$Fe scintillation peak from the bottom-most portion of the $^4$He target region.  This bottom-most portion of the $^4$He is particularly high in rate (see Fig.~\ref{fig:delay_times}) which we hypothesize could be due Auger electron emission by the source.  We adjust the energy scales of each dataset by at most 6\% to counter observed variation in the median of this scintillation peak from this region of $^4$He.

The result of the calibration effort is an estimate of energy deposited into the athermal phonon system of the Si: ``\evSi''.

\section{Detector Performance Studies}

\subsection{Evaporation Pulse Shapes}\label{sec:evap_shapes}

While a single ``typical'' evaporation pulse shape was chosen for the OF analysis, the observed evaporation pulse shapes in Fig.~\ref{fig:delay_shapes} show small but important variations.  The evaporation shapes are estimated by taking the residual of the prompt template fit and iteratively determining the median at each time bin. 
 The traces used in this procedure pass the ``standard quality'', ``in-$^4$He'', and ``$^{55}$Fe'' cuts.  The traces in Fig.~\ref{fig:delay_shapes} are the time series of these iteratively-determined-medians.  The upper panel shows the shape as a function of fill height, while the lower panel shows the shape as a function of interaction depth and includes a 100 kHz low-pass filter to account for the lower number of traces used in constructing the median.  The evaporation signal is broader when the fill height is low, indicating a greater dispersion in the evaporated atom arrival time.  This is a result of the larger vacuum gap and the signal timing being dominated by the atomic propagation in the vacuum: compared to the quasiparticle propagation the atoms are slower (by roughly a factor of 2~\cite{hertelDirectDetectionSubGeV2019}), and have greater velocity and angle (path length) dispersion, which each act to extend the pulse duration.

Pulse shape variation is also seen in the falling edge when comparing evaporation waveforms from different interaction depths.  On this falling edge we identify the presence of a number of features. One hypothesis is that these could be ``quasiparticle echo'' features, where quasiparticles could be reflecting from the bottom and sides of the target volume.  Specifically, in the 27.7~mm fill state shown in the lower panel of Fig.~\ref{fig:delay_shapes}, events near the top of that volume appear to show a small feature at $\sim$0.4~ms.  This would match the expectation in timing for an echo off the bottom of the target region, given a two-way quasiparticle travel distance of $\sim$50~mm and a quasiparticle group velocity of $\mathcal{O}$(200~m/s)~\cite{hertelDirectDetectionSubGeV2019}.

\subsection{Scintillation and Evaporation Amplitudes}\label{sec:iron_blobs}

\begin{figure*}[htbp]
    \begin{center}
        \includegraphics[width=0.6\textwidth]{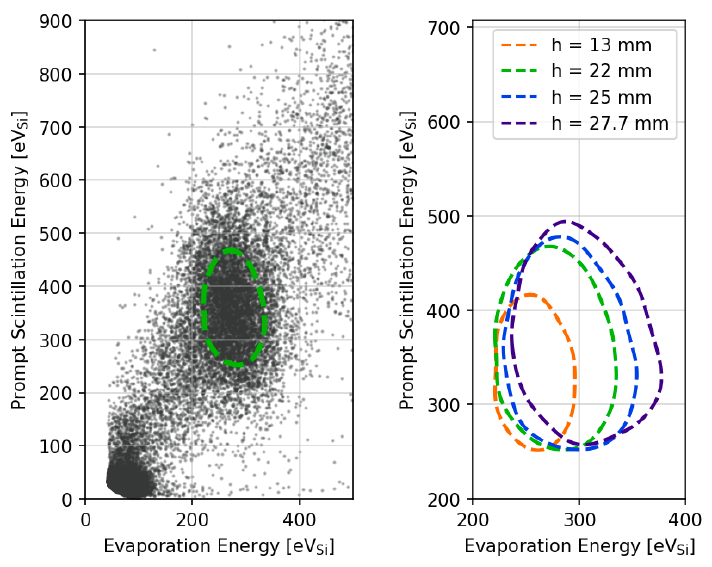}
    \end{center}
    \caption{Distributions of prompt-scintillation versus evaporation signal amplitudes.  The left panel shows data from the 22~mm fill height.  The 5.9~keV $^{55}$Fe peak is highlighted.  The right panel shows $^{55}$Fe contours in the plane for multiple $^4$He fill heights.  See Section~\ref{sec:iron_blobs} for discussion.}
    \label{fig:amplitude_relationships}
\end{figure*}

In this and the following subsections, we employ the two-template optimal filter processing described in Sec.~\ref{sec:processing} to highlight observations regarding the evaporation and prompt scintillation amplitudes, and the delay times between the two signals.

After each waveform is processed to extract amplitude estimates of the evaporation and prompt scintillation components individually, the two amplitudes can be plotted on an event-by-event basis, as in Fig.~\ref{fig:amplitude_relationships} (left panel). In this analysis (excluding the green contour, cuts for which are described below), we apply the ``standard quality'', ``22~mm evaporation time delay'', and ``$\Delta\chi^2$'' cuts.  This left panel is specifically for the \textit{h}~=~22~mm fill state.  The 5.9~keV calibration peak appears well above threshold in both signal channels.  In prompt scintillation, the peak appears as $\sim$350~\evSi~of energy in the Si. We expect the energy of an electron recoil signature in $^4$He to be partitioned roughly equally among singlet excimers (scintillation), triplet excimers, and quasiparticles.  More precisely, for a 5.9~keV deposit we expect $\sim$1875~eV of energy in the singlet channel, the dominant producer of prompt scintillation~\cite{collaborationScintillationYieldElectronic2022}. The observed $\sim$350~\evSi~scintillation signal amplitude is consistent with that $\sim$1875~eV expectation given the solid angle of the calorimeter (between 15\% and 40\%).  This work is not yet of sufficient precision to comment on contributions to the prompt signal from either prompt triplet deexcitation (via Penning ionization) or IR photons from higher excited states.  These topics will be the focus of future investigations.

The 1.5~keV calibration peak is also observed in Fig.~\ref{fig:amplitude_relationships} (left panel), appearing near threshold with scintillation energy of $\lesssim$170~\evSi.  At 1.5~keV, the mean observed number of 15.5~eV scintillation photons will be $\sim$5, explaining the observed large variance on the 1.5~keV scintillation signal amplitude.

In the evaporation channel, we see the 5.9~keV energy deposit appears as $\sim$300~\evSi~of energy in the Si of the calorimeter.  Again understanding that electron recoil energy is partitioned roughly equally in $^4$He between singlet excimers, triplet excimers, and quasiparticles, we estimate this $\sim$300~\evSi~corresponds to $\sim$1975~eV of initial quasiparticle energy, or a quasiparticle channel gain of $0.15 $~$\pm$~$ 0.01$, where the uncertainty comes from combining the calibration uncertainty with the evaporation pulse shape mismatch (variation) systematic in quadrature, and we do not include an uncertainty on the model.  This evaporation channel gain combines multiple effects, in particular the evaporation efficiency (fraction of $^4$He quasiparticles which lead to evaporation) and the van der Waals gain (energy per $^4$He atom adsorbed onto the Si surface).  While it is difficult to separately measure these two effects, if we assume the initial quasiparticle energy is approximately 1~meV and a van der Waals gain of 10~meV/atom, this corresponds to a quasiparticle channel gain of approximately 10$\times$.  Therefore, approximately 300~\evSi/10 = 30 \evHe worth of quasiparticles evaporated, implying an evaporation efficiency of 1-2\%.  This small evaporation efficiency emphasizes the importance of van der Waals gain to the technology's viability: the technology takes a small number of evaporated atoms and converts those atoms to a sizeable calorimetric signal.

It is also instructive to study the evaporation and prompt-scintillation amplitudes as the $^4$He fill height is varied. This variation is shown for the $^{55}$Fe peak in Fig.~\ref{fig:amplitude_relationships} (right panel), where we construct event-density contours after applying the ``standard quality'', ``in-$^4$He'', and ``$^{55}$Fe'' cuts.  The event density is normalized by live-time at each fill state, and for each fill state a contour is constructed at 0.0125~counts/(eV$_{\mathrm{Si}}^2$-hour).  We see that as the fill level increases, the mean signal amplitudes (in both channels) increase due to solid angle effects.  The evaporation signal evolves with fill state differently.  Because evaporation is dominated by a narrow cone of quasiparticles close to normal incidence with the liquid surface ($\lesssim$17$^\circ$)~\cite{bandlerParticleDetectionEvaporation1992,simon_bandler_detection_1994}, a roughly depth-independent number of atoms evaporate.  This said, atoms then evaporate from the surface with a full 2$\pi$ range of outgoing solid angle.  This means the transport of energy to the sensor through the atomic state is highly sensitive to the geometry and thickness of the vacuum gap, and can play an important role in evaporation signal gain even for events near the bottom surface.

\subsection{Time Delay Distributions and Interaction Depth}\label{sec:time_delay}

\begin{figure}[tbp]
    \begin{center}
        \includegraphics[width=\columnwidth]{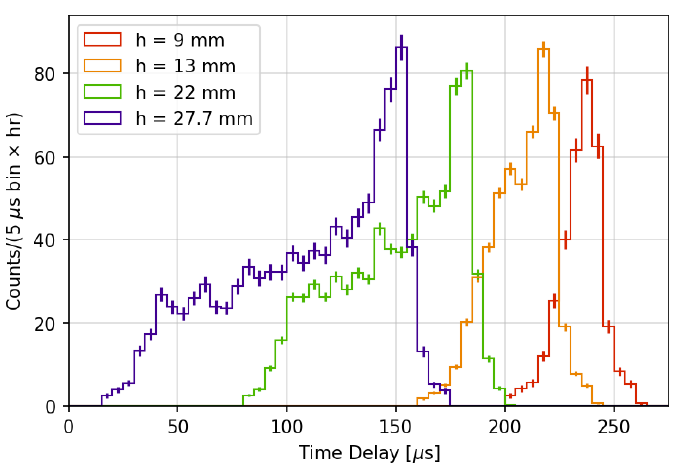}
    \end{center}
    \caption{Evaporation delay time distributions for $^{55}$Fe events.  The delay time depends on interaction depth, with the bottom of the detector appearing at the longest delay times (of each fill state).  See Section~\ref{sec:time_delay} for discussion.}
    \label{fig:delay_times}
\end{figure}

The distribution of evaporation signal delay times can be studied to better understand the physics of quasiparticle and atomic propagation.  This delay time is shown in Fig.~\ref{fig:delay_times} for $^{55}$Fe events with a variety of fill heights, $h$.  Here we apply the ``standard quality'', ``in-$^4$He'', and ``$^{55}$Fe'' cuts, as well as requiring the two-pulse-fit $\chi^2$ to be less than $10^4$.  At each fill height an overdensity is evident at the longest delay times, representing the bottom of the $^4$He target region.  One hypothesis is that this feature could be due to Auger electrons from the source, which exhibit a dramatically shorter path-length in the $^4$He.  

At the lowest fill levels the time delay is dominated by atomic propagation through the vacuum state, and we can estimate that atomic velocity scale as v$_{\mathrm{atom}}$~=~$\sim$30~mm/250~$\mu$s~=~$\sim$100~m/s.  At higher fill levels the quasiparticle state accounts for more of the path length.  As the quasiparticle state is faster than the atomic state, this shifts delay time distributions to smaller values.  Additionally, as more interaction depths become possible the time delay distribution widens.  To estimate the quasiparticle velocity we consider the state where quasiparticles make up the most path length: \textit{h}~=~27.7~mm.  In this state the top of the liquid appears at a delay time of $\sim$40~$\mu$s, and the bottom appears at $\sim$150~$\mu$s, implying a quasiparticle velocity of v$_{\mathrm{qp}}$~=~$\sim$27.7~mm/110~$\mu$s~=~$\sim$250~m/s.  The $^4$He quasiparticle velocity is highly momentum dependent~\cite{hertelDirectDetectionSubGeV2019}, and this observed velocity implies that the evaporation process is dominated by the fastest quasiparticle states (R$^+$ rotons), consistent with expectation~\cite{hertelDirectDetectionSubGeV2019}.

\subsection{Amplitude Variations with Depth}\label{sec:amp_vs_depth}

\begin{figure*}[tb]
    \begin{center}
        \includegraphics[width=\textwidth]{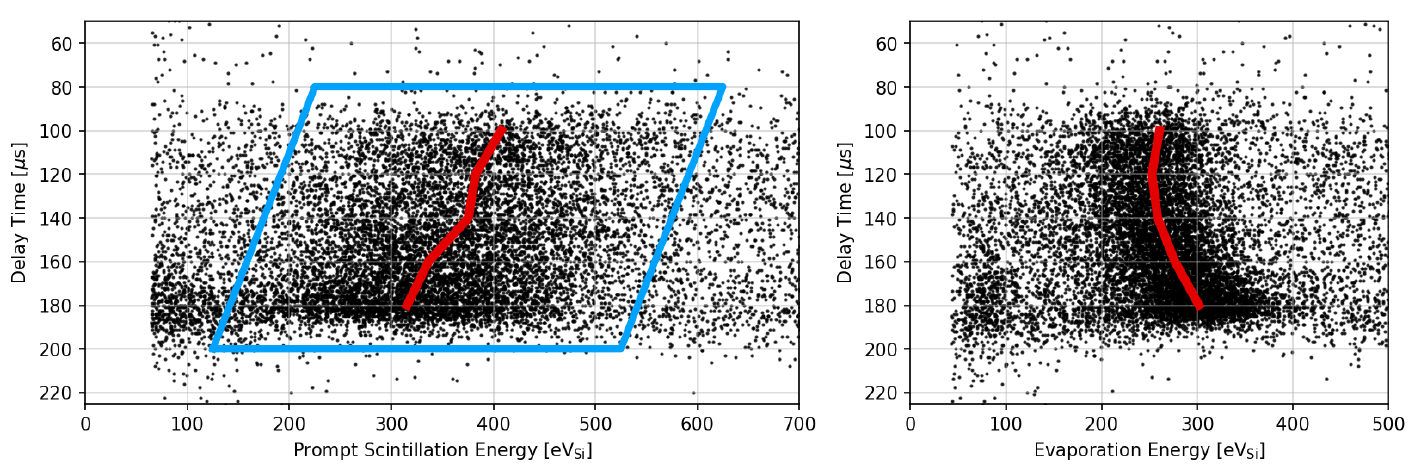}
    \end{center}
    \caption{Scintillation (left) and evaporation (right) signal amplitudes as a function of evaporation delay time, a proxy for event depth within the $^4$He.  Figure discussion in Section~\ref{sec:amp_vs_depth}.  Fill height for this data was \textit{h}~=~22~mm.  The main band results from 5.9~keV $^{55}$Fe X-rays.  The 1.5~keV aluminum fluorescence can be seen near threshold and at the bottom of the active region (bottom left of the plots).  Red lines indicate the peak of the $^{55}$Fe population, measured with a Gaussian kernel density estimator in 20~$\mu$s bins.  The blue lines indicate the trapezoidal ``$^{55}$Fe'' cut referenced in Section~\ref{sec:data_selection}.}
    \label{fig:amplitudes_vs_delay}
\end{figure*}

We can combine the amplitude analysis with the event depth information gathered from delay time to understand how signal efficiency depends on event depth within the $^4$He.  Figure~\ref{fig:amplitudes_vs_delay} shows the evaporation and prompt scintillation energies as functions of delay time for the \textit{h}~=~22~mm fill state.  Note that the delay time has been inverted to afford a more intuitive orientation: the bottom of the $^4$He volume is at the bottom of the plot.  In this analysis we apply the ``standard quality'' and $\Delta\chi^2$ cuts, and well as require a scintillation pulse amplitude greater than 65~\evSi.  The 5.9~keV population is clearly visible at all delay values (all depths), while the near-threshold 1.5~keV population is visible only very close to the X-ray source itself (due to its short mean-free-path), near the bottom of the $^4$He volume.

The variation of the 5.9~keV signal with depth provides further insight into the physics of the two channels.  In the right panel of Fig.~\ref{fig:amplitudes_vs_delay} the scintillation signal is seen to decrease with distance from the calorimeter, matching expectation from decreasing sensor solid angle.  In the left panel the opposite trend with depth appears in the evaporation signal:  evaporation signals are larger for interactions occurring deeper within the $^4$He.  This counter-intuitive position dependence can be explained by considering those quasiparticles which are emitted from the interaction site but are initially outside the narrow evaporation critical angle previously mentioned.  If this much larger quasiparticle population has some probability for diffuse reflection at interfaces, then these diffusely reflected quasiparticles have some probability to then fall within the evaporation critical angle and contribute to the signal.  The evaporation signal can thus be broken into two contributions: a `primary' contribution made of of quasiparticles which propagate directly to the liquid surface within the critical angle, and a `secondary' contribution made of quasiparticles which evaporate after one or more reflections.  The primary contribution will be nearly depth-independent.  The secondary contribution will increase in amplitude with depth, as the combined solid angle of the bottom and side walls is larger for deeper positions. The scale of the effect in the left panel of Fig.~\ref{fig:amplitudes_vs_delay} is consistent with a quasiparticle reflection probability at metal surfaces of $\sim$0.3, consistent with previous literature~\cite{bandlerParticleDetectionEvaporation1992}.  The `primary' contribution would then represent a depth-independent $\sim$200~\evSi~signal and the `secondary' contribution would represent a boost of $\sim$20\% at the top and $\sim$40\% at the bottom.  As mentioned in Sec.~\ref{sec:evap_shapes}, the presence of reflection quasiparticles is further supported by the ``echo'' features in Fig.~\ref{fig:delay_shapes}.  Additionally, we highlight a subtle deviation from this ``decreasing trend" at the top of the detector (100-120~$\mu$s). This slight increase in evaporation signal for energy deposits near the liquid surface may result from simple geometric considerations: for events near the liquid surface, the evaporation is from a region of small radial extent, meaning fewer of the evaporated atoms will be at large radius where sensor collection efficiency is low.

\subsection{Triplet Signals}\label{sec:triplets}

\begin{figure*}[tbp]
    \begin{center}
        \includegraphics[width=\textwidth]{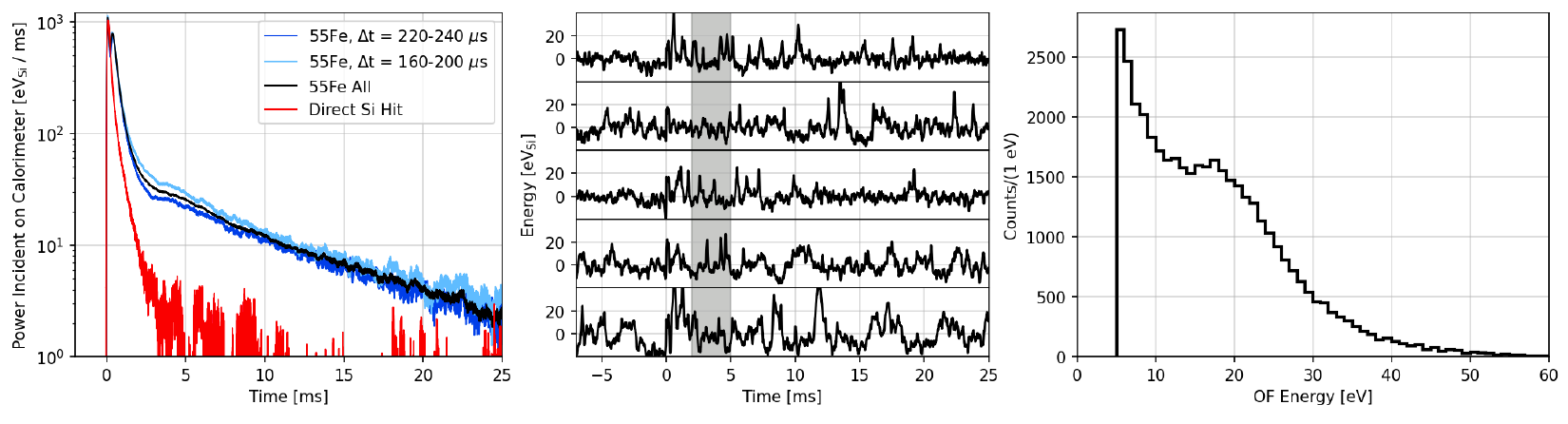}
    \end{center}
    \caption{Observations of triplet excimer decays at late times, using the \textit{h}~=~13~mm data set.  \textit{Left:} Waveforms averaged over many events, see Section~\ref{sec:triplets} for further details.  \textit{Center:} Individual residual waveforms from the $^{55}$Fe peak after subtracting the prompt scintillation and evaporation components, revealing small features consistent with individual triplet deexcitations.  The grey region indicates the histogram integration window for the right panel.  \textit{Right:} A spectrum of the amplitude of the individual small peaks shown in the center panel, indicating rough consistency with the expected 15.5~eV energy.}
    \label{fig:triplets}
\end{figure*}

As visible in Fig.~\ref{fig:pulses}, times following the evaporation signal are dominated by the deexcitation of long-lived triplet excimers.  At high isotopic purity and T$<$100~mK, the long-lived triplet excimers propagate ballistically from the interaction site to the $^4$He surfaces. Assuming a velocity of 1-5~m/s~\cite{zmeevObservationCrossoverBallistic2013}, we expect propagation times of $\lesssim$10~ms in the few-cm $^4$He region of this work.  Upon arrival at the bottom and sides of the $^4$He region, the triplet excimers are expected to promptly decay through electron exchange with the metal~\cite{carter_calorimetric_2017}.  If the triplet excimers instead arrive at the liquid surface, they are expected to diffuse along the surface in a trapped state (\cite{sethumadhavan_electrical_2006} and references within), until deexcitation can proceed through interaction with another triplet excimer (Penning ionization), a $^3$He impurity, or through diffusion all the way to the metal side wall.

Observations of triplet excimer decays are shown in Fig.~\ref{fig:triplets}.  In all panels the ``standard quality'' cuts are applied.  The left panel shows averaged waveforms in a long time window, up to 25~ms following the prompt scintillation signal.  To facilitate the averaging, traces were required to have a start time greater than $200~\mu$s, which due to our triggering algorithm removes events with small scintillation amplitudes.  The red trace selects for events occurring directly in the sensor by requiring that both $100 < \Delta\chi^2 < 500$ and the scintillation amplitude was between 300 and 400~nA.  The remaining curves instead select for events in the helium by applying the `in-$^4$He'' and ``$^{55}$Fe'' cuts, in addition to the evaporation delay time cuts listed in the plot legend.  The y-axis unit of \evSi/ms is scaled from TES current per time using the same calibration factors found previously.  It can be seen that, relative to an event occurring in the silicon itself (the red ``Direct Si Hit'' average), $^4$He events exhibit evaporation and then additional signals at later times.  At t=$\sim$4~ms, a ``shoulder'' appears with depth-dependent amplitude.  This few-ms timescale is consistent with that expected for the prompt quenching of excimers directly incident on metal surfaces given the cm-scale cell dimensions and m/s-scale triplet velocities.  At later times, the triplet deexcitation rate follows an exponential decay with $\tau \approx 5.4$~ms.  This time period may represent the deexcitation of triplets on the liquid surface, and the exponential nature of the process may indicate triplet-triplet deexcitation (for which a t$^{-1}$ dependence would be expected) is strongly subdominant to other deexcitation mechanisms.  Crucially, this demonstrates that triplets are removed from the detector system on timescales much faster than their 13 s lifetime~\cite{mckinseyRadiativeDecayMetastable1999}.

The average energy received by the calorimeter during the triplet-dominated late time window can be found for 5.9~keV deposits by integrating the average waveforms of Fig.~\ref{fig:triplets} (left).  An integral of the 4~ms~-~25~ms window corresponds to $\sim$200~\evSi~of received energy.  When compared to the $\sim$320~\evSi~prompt scintillation signal at this same \textit{h}~=~13~mm fill state, the triplet deexcitation signal exhibits $\sim$60\% as much energy received. As discussed in Section~\ref{sec:detector_main}, the expected triplet excitation yield in $^4$He is lower than the singlet excitation yield by a factor of $\sim$0.75.  After accounting for this difference in yield, the data can be interpreted as implying that the singlet channel ``gain'' (energy observed per energy excited) is only incrementally higher than the corresponding triplet channel gain.

The middle panel of Fig.~\ref{fig:triplets}, shows several typical waveforms from the $^{55}$Fe peak after a procedure to highlight triplet deexcitation features, selecting from events passing the ``standard quality'', ``in-$^4$He'', and ``$^{55}$Fe'' cuts.  To construct this version of the event waveforms, the fitted scintillation and evaporation pulses (resulting from the standard two-template OF procedure) are subtracted, leaving a roughly flat residual.  This residual waveform is then passed through the scintillation template OF, including the appropriate calibration factor to convert OF amplitude to eV$_{\mathrm{Si}}$.  In this filtered and energy-calibrated version of the residual waveforms, a number of sharp peaks consistent with individual photon arrivals are evident in the period following t=0, and the energy of these peaks is seen to be at the expected $\mathcal{O}$(10~eV$_{\mathrm{Si}}$) scale.

The right panel of Fig.~\ref{fig:triplets} shows a histogram of such peak amplitudes.  After many 5.9~keV waveforms are passed through the calibrated optimal filtering procedure of the middle panel, the height of all peaks greater than 5 eV with a width of 100 samples ($80~\mu$s) in a \textit{t}~=~2-5~ms window are tabulated and histogrammed.  Noise fluctuations with a standard deviation of 3.2~\evSi~contribute significantly to population of peaks in this region, and also contribute to a broadening of the observed feature.  Gaussian fits have been performed and well describe this peak, but the fitted mean is highly sensitive to the assumed shape of the underlying background and so is not reported.  The relatively high value of this $\approx$15~\evSi peak on the few-ms time-scale (together with the overall high triplet excimer channel gain factor mentioned earlier) may imply that triplet excimer deexcitations on metal surfaces are sharing only a small fraction of their energy with the copper surfaces, appearing instead as photons of nearly the full 15.5~eV energy.

\section{Implications for Dark Matter Sensitivity}

The measured evaporation channel gain of $0.15$~$ \pm $~$0.01$ allows us to better project future sensitivity of this technology to dark matter detection via nuclear recoils.

We first point out that even at the current R\&D stage, we are entering the ``quasiparticle-only'' signal regime at low energies, where the fraction of recoil energy appearing as electronic excitation is either negligible or truly zero.  For this present data, the 5 sigma detection threshold for energy in the $^4$He quasiparticle system is approximately 145~eV, estimated as follows.  First, the optimal filter provides the 5 sigma threshold, in current, of the evaporation pulse shape.  This is then scaled to \evSi~with the energy calibration measured for the scintillation pulse shape, scaled by the relative areas of the pulse templates (this assumes the sensor is linear).  Finally, the threshold in \evSi~is divided by the 0.15 evaporation channel gain to get the threshold in \evHe.  A 145~eV nuclear recoil has some chance of producing a single excimer (singlet or triplet)~\cite{hertelDirectDetectionSubGeV2019, collaborationScintillationYieldElectronic2022}, but most nuclear recoils at this energy will convert all recoil energy into a population of quasiparticles of equal total energy.  Further, the momentum distribution of quasiparticles is expected to vary only weakly with recoil energy~\cite{youSignaturesDetectionProspects2022}, and so expect a similar small variation in evaporation channel gain with recoil energy.

An evaporation channel gain of less than 1 means the recoil energy threshold in $^4$He is currently weaker than the threshold in the calorimeter itself.  However, the lower-mass $^4$He nucleus (A~=~4) boosts the DM recoil energy in $^4$He relative to Si (A~=~28) by a factor of 28~/~4~=~7, coincidentally countering the low evaporation channel gain:   $0.15\times7=1.05$.  More succinctly: for the observed quantum evaporation gain of 0.15, the DM mass threshold in the $^4$He and the Si are nearly equal.

A nuclear recoil energy threshold can be translated to a simple estimate for DM mass threshold by assuming a DM velocity equal to the conventional escape velocity of $v_{esc}$~=~544~km/s plus the local standard of rest velocity $v_0$~=~238~km/s~\cite{baxterRecommendedConventionsReporting2021}.  The current 145~eV energy threshold then implies a sensitivity to dark matter candidates with $m_\mathrm{DM}\gtrsim$~220~MeV/c$^2$.

In future work, the calorimeter energy threshold can be significantly reduced.  With a heat-free film-stopping method demonstrated, the TES transition temperature can be reduced to less than 20~mK, and such devices have already been fabricated and demonstrated within the SPICE/HeRALD collaboration.  Calorimeter energy thresholds of less than 1~eV are expected.

It is also reasonable to expect an increase in the evaporation gain factor in the near term.  First, the van der Waals gain (energy per adsorbed atom) can be increased by moving away from Si to alternative calorimeter substrates which more strongly polarize the $^4$He atom upon arrival.  For example, an Al$_2$O$_3$ (sapphire) substrate is expected to increase the van der Waals adsorption energy by a factor of $\sim$2~\cite{zhangFirstPrinciplesInvestigation2016}, and successful device fabrication on Al$_2$O$_3$ has already been demonstrated within the SPICE/HeRALD collaboration.  Second, while this current work demonstrates some small increase in gain from quasiparticle reflection at material interfaces, a moderate increase in quasiparticle reflection probability (through alternative coatings of detector surfaces) would result in a faster-than-linear increase in the evaporation probability after multiple reflections.

For illustrative purposes, we project the sensitivity of a near-term HeRALD experiment with moderate improvement in each aspect.  If one assumes a calorimeter energy threshold of $\sim$1~eV, a $\sim$2$\times$ increase in van der Waals gain, and a $\sim$2$\times$ increase in evaporation efficiency (a combined gain factor of $0.15\times2\times2=0.6$), then the resulting a $^4$He nuclear recoil energy threshold would be $\sim$1~eV$/0.6=~\sim$1.7~eV and the dark matter mass threshold would be $m_\mathrm{DM}\gtrsim$~30~MeV/c$^2$.

\section{Conclusions}

We have presented the design and first results of a prototype $^4$He detector, `HeRALD v0.1'.  The main goals of this R\&D phase were to determine the efficacy of the cesium film-stopping system and to make first observations of particle interactions in the superfluid $^4$He target.  These studies show that the cesium film-stopping system can be made practical, and that it is effective over the course of a two-month operation.

The newly-measured evaporation channel gain of $0.15 \pm 0.01$ is sufficient for interesting DM sensitivity in the near-term after transitioning to lower-energy-threshold calorimetry.  The existing gain factor results in DM mass thresholds similar to the Si calorimeter itself.  Future improvements to this gain factor are expected through increased van der Waals adsorption energies and increased $^4$He quasiparticle reflection probabilities.

We also highlight the importance of the newly observed triplet excimer deexcitation timescale.  The long triplet excimer lifetime in the bulk has implied that these excimers could potentially form a background of eV-scale `dark counts' in a future DM search.  The newly observed high triplet signal gain and few-ms exponential decay timescale suggests that these excimers are largely removed on short timescales.

This is only a first demonstration of the HeRALD architecture.  We anticipate a significant reduction in energy threshold, giving HeRALD a path to probing DM masses at or below 30~MeV/c$^2$ within a small number of years. 

\begin{acknowledgments}

We would like to thank Maryvonne De Jesus for providing the $^{55}$Fe source used in these studies.  Additionally, we would like to thank George Seidel for useful conversations throughout the course of this work.  H.D.P would like to thank the hospitality of the MANOIR group at the IP2I in Lyon where a portion of this analysis was performed during a stay funded in part by the Chateaubriand Fellowship.  This work was supported in part by DOE Grants DE-SC0019319 and DE-SC0022354, and DOE Quantum Information Science Enabled Discovery (QuantISED) for High Energy Physics (KA2401032). This material is based upon work supported by the National Science Foundation Graduate Research Fellowship under Grant No. DGE 1106400. This material is based upon work supported by the Department of Energy National Nuclear Security Administration through the Nuclear Science and Security Consortium under Award Number(s) DE-NA0003180 and/or DE-NA0000979.  Work at Argonne is supported by the U.S. DOE, Office of High Energy Physics, under Contract No. DE-AC02-06CH11357.  Work at Lawrence Berkeley National Laboratory is supported by the U.S. DOE, Office of High Energy Physics, under Contract No. DE-AC02-05CH11231.  W.G. and Y.Q. acknowledge the support by the National High Magnetic Field Laboratory at Florida State University, which is supported by the National Science Foundation Cooperative Agreement No. DMR-2128556 and the state of Florida.

\end{acknowledgments}

\appendix

\section{Cryostat Systems}\label{sec:cryostat}

\begin{figure}[tbp]
    \begin{center}
        \includegraphics[width=0.9\columnwidth]{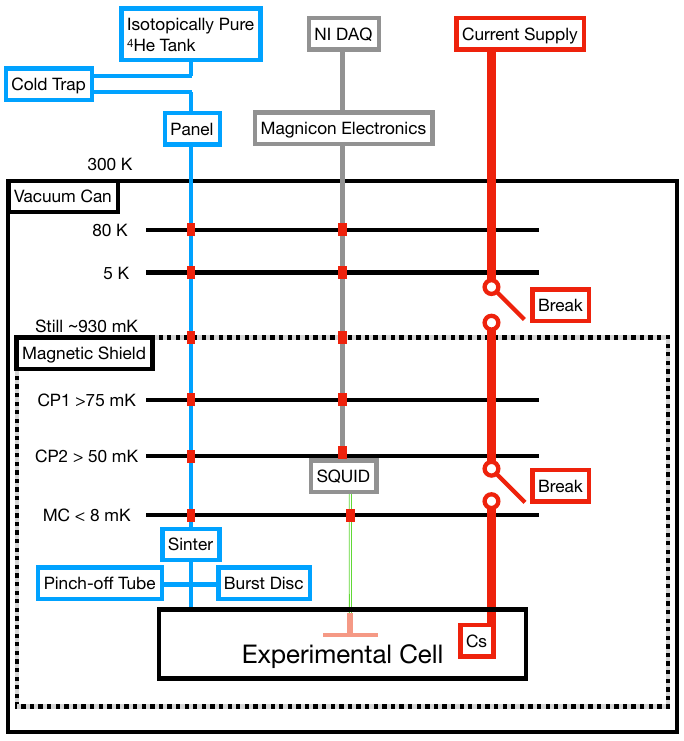}
    \end{center}
    \caption{Schematic of experimental subsystems within the dilution refrigerator.  Shown in blue is the isotopically-pure $^4$He injection system.  The Magnicon SQUID readout is shown in grey, the sensor is shown in orange, and the NbTi twisted-pair connections between the SQUID and sensor are shown in pink.  The $^4$He and readout systems are heat-sunk at each thermal stage, represented by red blocks in-line with the readout.  Finally, the cesium dispensing system is shown as a thick red line with switches representing the post-evaporation break points just above the still and mixing chamber (MC) stages.  Thermal shields at 80~K and 5~K are not shown.  The shield at the Still is made of Cryoperm and doubles as a magnetic shield.}
    \label{fridge_diagram}
\end{figure}

The dilution refrigerator used for this study is a CryoConcept Hexadry UQT-B 400.  Figure~\ref{fridge_diagram} schematically illustrates the routing of several components through the cryostat to the cell: a capillary for both filling and removing $^4$He, high-current leads for cesium evaporation, and a DC SQUID readout chain for the sensor.

\subsection{$^4$He Handling System}

The capillary system carries $^4$He from room temperature storage to the experimental cell.  The $^4$He itself was isotopically enriched at Lancaster University\footnote{\url{https://www.lancasterhelium.uk/}} with a $^3$He concentration below $5\times10^{-13}$~\cite{hendryContinuousFlowApparatus1987}.  At room temperature the $^4$He is kept in a large low-pressure ($\sim$1~atm) storage vessel.  A gas handling system can circulate the $^4$He through a liquid-nitrogen cooled activated-charcoal cold-trap.  During filling a mass flow controller (MFC) is used to regulate the $^4$He condensation.  To preserve $^4$He purity, all connections are either welded or VCR (metal gasket).  The only exception to this design philosophy was a ceramic electrical break near the refrigerator flange which was mated using Swagelok compression fittings.  

The $^4$He connection between room-temperature and the 80~K stage of the refrigerator was made with a 1/4" VCR-flanged flexible bellows.  This bellows accommodates thermal contractions during the cooldown and provides moderate surface area for any impurities which escape the cold-trap.  Below the 80~K stage the $^4$He is carried through a stainless steel capillary (2.1~mm OD, 1.7~mm ID).  At each temperature stage, including 80~K, this line is heat-sunk by breaking the capillary and soldering the ends into a copper block with a large (4~mm ID) bore.  The $^4$He line connects to the cell through a 0.5~$\mu$m sinter\footnote{Swagelok SS-4-VCR-2-.5M}, intended both to dampen any thermo-acoustic oscillations during filling and also to prevent any flow of gas atoms from the warmer stages to the cell interior during data-taking.  The cell and capillary are typically pumped out via a copper pinch-off tube mounted at the top of the cell (however, the pinch-off tube was not used in this experiment and the cell was only evacuated through the capillary).  In case of a capillary blockage or a sudden temperature rise, the cell is fitted with a burst disc and the fridge vacuum volume is fitted with with a pressure relief valve.

\subsection{Film Stopping and the Cesium System}

The HeRALD collaboration considered two techniques of film stopping: the knife-edge and cesium methods.

The knife-edge method, demonstrated at higher temperatures (particularly by X-ray observing satellites~\cite{ezoePorousPlugPhase2017,ishikawaPorousPlugSuperfluid2010,shirronSuppressionSuperfluidFilm1998}), interrupts film flow using an extremely sharp corner separating the bulk $^4$He and the film-free region. The van der Waals attraction of the $^4$He to the knife edge material forces the film to thin in the vicinity of the edge.  For a sufficiently sharp edge, the film can be thinned below its critical thickness such that the film will no longer exhibit superflow.  However, this critical film thickness decreases with temperature such that superflow can appear in films as thin as 2.1 atomic layers~\cite{chanSuperfluidityThin41974,scholtzThirdSoundHealing1974,brewerGaplessSurfaceExcitations1965} at $\sim$10~mK, HeRALD's operating temperature.  This nearly-atomic critical film thickness requires a near-atomically-sharp knife edge (a radius of curvature at or below the nanometer scale~\cite{mcclintockThicknessHeliumFilm1973,ishikawaPorousPlugSuperfluid2010,shirronSuppressionSuperfluidFilm1998}).  We investigated fabrication of such devices through anisotropic etching of Si, and achieved knife edge radii only as low as $\sim$7~nm.  Given the challenges of this technique and the promise of the cesium method, the knife edge development was discontinued.

The mechanism of the cesium method was described in Sec.~
\ref{sec:film_stopping}.  The evaporation system used to implement this method must overcome several technical challenges stemming from cesium's extreme reactivity and high vapor pressure~\cite{oswaldkubaschewskiMetallurgicalThermochemistry1967}.  The substrate for cesium deposition is gold-plated to reduce that substrate's cesium reactivity.  To avoid oxidation, the cesium is evaporated directly onto the desired surfaces within the cell, after it is mounted on the mixing chamber, and after cooling the mixing chamber stage and cell to $\sim$4~K.  Cesium is evaporated from the dispensers towards the gold-plated substrate, and copper baffling prevents cesium from entering the experimental area below.  The required 7.5~A evaporation current is delivered via $\sim$1~mm diameter Kapton-insulated copper wire.  This cabling would represent a significant heat load to the coldest stages of the refrigerator, and so this heat load is eliminated using ``breaks'' in the current leads just above the mixing chamber and still stages.  At these breaks, the copper wire is pressed between copper (grade 101) spring clamps during the deposition.  At room temperature a linear direct mechanical feed-through is manipulated after cesium deposition in order to open the two thermal breaks.  This manipulation also engages an electrical grounding spring to the cesium leads at the mixing chamber stage to mitigate electrical noise pickup.

\subsection{Detector Readout}

The final detector system is a DC SQUID from Magnicon for TES readout.  The TES array is connected to the 16-SQUID series array using a superconducting NbTi wire twisted pair, in parallel with a 2~m$\Omega$ shunt resistor patterned onto the Magnicon SQUID chip itself.  The SQUID is shielded magnetically by a superconducting Nb can around the array, and the entire low-temperature readout is surrounded by a magnetic shield which doubles as our still stage thermal shield.  The SQUID array is controlled at room temperature with an XXF-1 electronics system.  Signals are filtered with a 500~kHz low-pass anti-aliasing filter, and recorded at 1.25~MHz using a National Instruments \mbox{PCI-e 6376} data acquisition card controlled by the \texttt{polaris} package~\cite{suerfuPolarisGeneralpurposeModular2018}, wrapped within \texttt{pytesdaq}~\cite{Pytesdaq2022}.  

\section{Detector Preparation}\label{sec:detector_preparation}

Prior to the data-taking, we first prepare the cell and cesium film.  This begins at room temperature by evacuating the cell to at least $\sim$10$^{-4}$~mbar.  While continuing to pump we outgas the cesium dispensers (ramping up to 3~A, in 10-minute steps of 1~A), raising the dispenser temperature to $\sim400~^{\circ}$C~\cite{p.dellaportaAlkaliMetalGeneration}.  We end the pumping after returning to base pressure.  Once the refrigerator is cooled to $\sim$4~K and we can ensure a cryogenic vacuum, we perform the cesium evaporation procedure (ramping up to 7.5~A, in 10-minute steps of 1~A).  After holding for 20 minutes at 7.5~A (a dispenser temperature of $\sim800~^{\circ}$C), the majority of cesium has been evaporated and the current is stopped.  This procedure heats the sensor platform to approximately 70~K.  Following the evaporation we disconnect the cesium leads with the breaks and mechanical feedthrough described in Appendix~\ref{sec:cryostat} and cool the refrigerator to base temperature.

\bibliography{june_12_2023}% Produces the bibliography via BibTeX.

\end{document}